\documentclass[twocolumn,secnumarabic,amssymb, nobibnotes, aps, prd]{revtex4-1}

\usepackage{subfig}
\usepackage{graphicx}
\usepackage{amsmath}

\begin{document}

\title[Scattering Cross Section in a Cylindrical]{Scattering Cross Section in a Cylindrical anisotropic layered metamaterial}%

\author{M. R. Forouzeshfard}%
\email[]{m.forouzeshfard@vru.ac.ir}
\affiliation{Department of Physics, Faculty of Science, Vali-e-Asr University, P. O. Box No: 77139-36417, Rafsanjan, Iran}
\author{Masoud Mohebbi}
\affiliation{Department of Physics, Faculty of Science, Vali-e-Asr University, P. O. Box No: 77139-36417, Rafsanjan, Iran}
\author{Aliyeh Mollaei}
\affiliation{Department of Physics, Faculty of Science, Vali-e-Asr University, P. O. Box No: 77139-36417, Rafsanjan, Iran}
\date{April 30, 2017}%

\begin{abstract}
To design a uniaxial anisotropic metamaterial a layered cylindrical metamaterial is introduced for TE polarization. 
Unlike to the previous work, which the layers were in radial direction, here the layers are in azimuthal direction.
Scattering efficiency for this metamaterial in different frequency is analyzed with solving Maxwell's wave equation.
It is observed that in some frequencies when the effective permittivity of the structure goes to zero the scattering efficiency would be negligible. This result approves the previous predictions. It is also found out that the scattering cancellation depends on the relative permittivity of the environmental medium for the cylinder. The finite element simulations are also confirmed the results.
\end{abstract}

\maketitle

\section{Introduction}\label{intro}
Cloaking and transparency are the subjects which attract many interests in all the world during the last decade \cite{pendry,ulf,forouzeshfard,alu}. One of the approach in cloaking is based on scattering suppression. Epsilon-Near-Zero (ENZ) material are very good candidate for this purpose \cite{alu,Irci,alu2,alu-ENZ}.

Introduction of hyperbolic layered metamaterials was a very significant progress to realization of uniaxial ENZ material \cite{hyperbolic1,hyperbolic2}. It was proved that the scattering cross section of a small arbitrary object at the center of a planar layered metamaterial with effective ENZ permittivity can be suppressed provided that dipole approximation be valid \cite{shalin}.

Kim's et. al. are also introduced a cylindrical layered metamaterial for both TE and TM polarization so that for a cylinder cavity made of alternating metal-dielectric structure in radial direction the invisibility condition in a specified frequency is happened when the effective permittivity of the structure goes to zero \cite{kim}. Another research also approved the result extracted by Kim's et. al., however it shows that the frequency of invisibility is also depend on the permittivity of the environment medium \cite{carlos}. Using graphene-coated nanowire in order to suppress the scattering cross section of a cylindrical cavity is also studied \cite{carlos-graphene}.

In this paper we also consider a cylindrical hyperbolic metamaterial, but with layered structure in azimuthal direction for TE polarization. The scattering efficiency of this structure is studied with solving Maxwell's wave equation analytically and the condition for minimizing the scattering cross section is found.

Some finite element simulations are also brought to approve the validation of the analytical solution.

This paper is organised as follow. In Sec.~\ref{analytic} we solve Maxwell's wave equation for a uniaxial anisotropic cylinder and the scattering cross section for this cylindrical scatterer is found analytically. In Sec.~\ref{ENZ} we proposed a hyperbolic layered metamaterial composed of metal dielectric layers and its effective permittivities is calculated using effective medium theory. Analysing the scattering cross section of this metamaterial structure and discussion on the condition of the invisibility is expressed in Sec.~\ref{result}.  We conclude the result in Conclusion section.

\section{analytical discussion}\label{analytic}
\begin{figure}[]
\center
\includegraphics[scale=0.3]{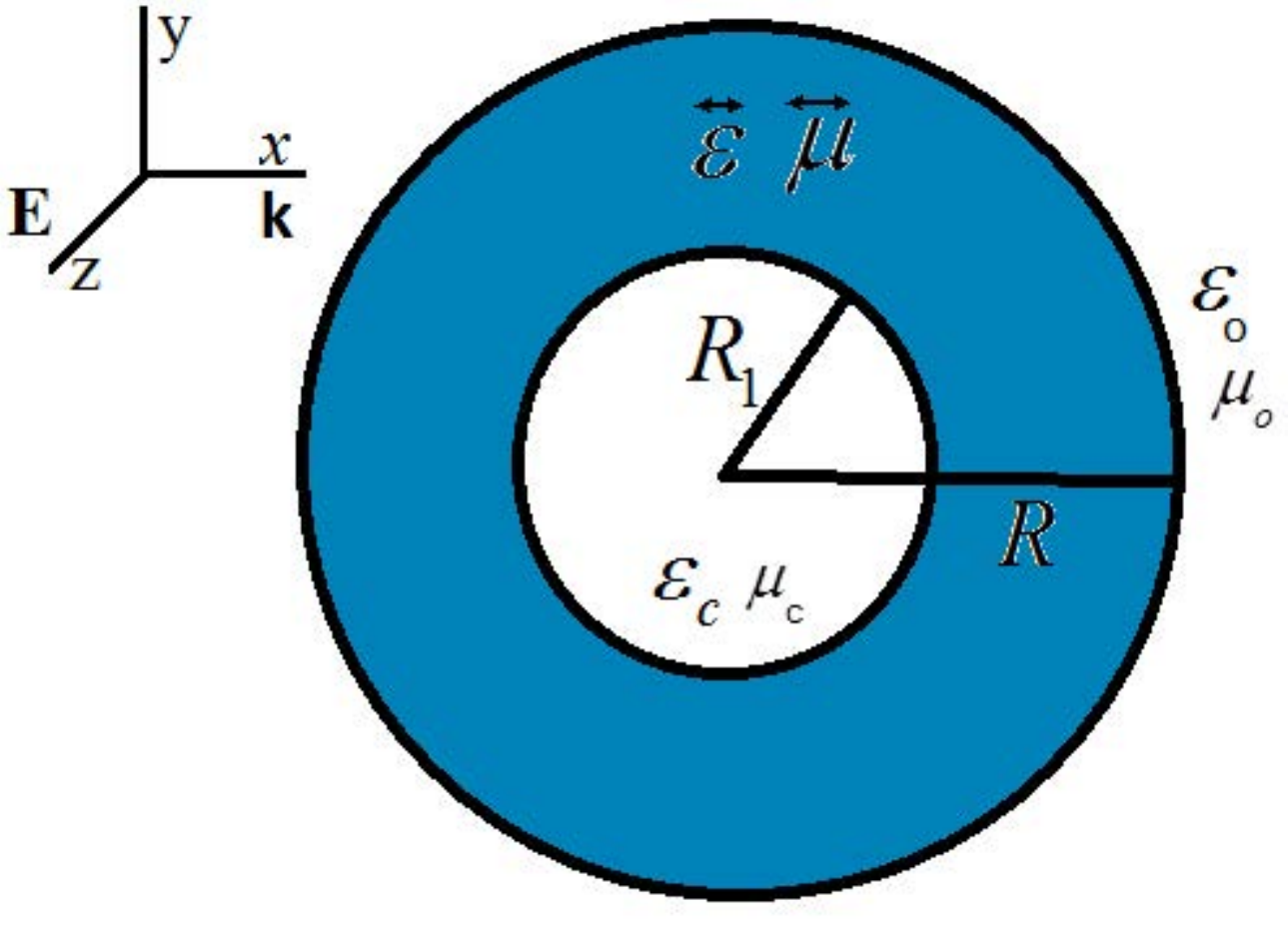}
\caption{A cylinder with anisotropic coated shell.}\label{cylinder}
\end{figure}

Consider a long coated cylinder with inner (outer) radius $R_1(R)$ so that the thickness of the shell is $T=R-R_1$. The electric permittivity and permeability for the core is $ \varepsilon_c $ and $\mu_c$, respectively and the shell is an anisotropic material with permittivity and permeability tensor expressed by 
\begin{eqnarray}\label{1q}
\overleftrightarrow{\varepsilon}  = \left( {\begin{array}{*{20}{c}}
{{\varepsilon _r}}&0&0\\
0&{{\varepsilon _t}}&0\\
0&0&{{\varepsilon _z}}
\end{array}} \right),\qquad\overleftrightarrow{\mu}  = \left( {\begin{array}{*{20}{c}}
{{\mu _r}}&0&0\\
0&{{\mu _t}}&0\\
0&0&{{\mu _z}}
\end{array}} \right).
\end{eqnarray}
The permittivity and permeability of the environment medium is denoted by $ \varepsilon_0 $, $ \mu_0 $ as shown in Fig.\ref{cylinder}.

An incident plane wave with TE polarization (electric field along $ z $ direction) is illuminated to the cylinder normally  so that the wave vector is in the $ x $-direction. 
We are going to find the constraint in which the total scattering cross section of this structure goes to zero. Lorentz-Mie scattering theory is used in order to find the solution of Maxwell's wave equation in all three regions. The approach which we apply to find the scattering efficiency is similar to the method in ref. \cite{Chen1} and \cite{Chen2} but for a TE plane wave illumination. 
Since for the outside region $ r>R $, we have an isotropic medium, so, for the electric field at the outside of the coated cylinder we have \cite{Bohren}
\begin{align}\label{2q}
{E_z}^{tot}={E_z}^{i}+{E_z}^{sc},
\end{align}
where in 
\begin{align}\label{3q}
{E_z}^{i}=E_0e^{ikx}={E_0}\sum_{n=-\infty}^{+\infty}{i^n}{J_{n}}\left({k_0{r}}\right)e^{in\phi},
\end{align}
is the electric field for the incident wave and $ E_z^{sc} $ is the scattered electric field from our structure and can be expanded in cylindrical coordinate as follow \cite{Bohren}
\begin{align}\label{4q}
{E_z^{sc}}={E_0}\sum_{n=-\infty}^{+\infty}{i^n}{a_n}{H_{n}^{(1)}}\left({kr}\right)e^{in\phi},
\end{align}
where $ H_n^{(1)} $ is the Hankel function of the first kind of order $ n $; $ k=k_0\sqrt{\varepsilon_0\mu_0} $ and $ k_0=\omega/c $.

Electric field in a core layer is as follow \cite{Bohren}
\begin{align}\label{5q}
{E_z}^{(1)}={E_0}\sum_{n=-\infty}^{+\infty}{i^n}{d_n}{J_{n}}\left({k_1{r}}\right)e^{in\phi},
\end{align}
where $ J_n(.) $ is a Bessel function of the first kind in order of $ n $ and $ k_1=k_0\sqrt{\varepsilon_c\mu_c}$.
Since the coated layer is an anisotropic medium, the solution of the Maxwell's wave equation in this region using the method similar to ref. \cite{Chen1} and \cite{Chen2} take the form as follow
\begin{align}\label{6q}
{E_z}^{(2)}={E_0}\sum_{n=-\infty}^{+\infty}{i^n}\left[{b_n}{J_{n'}}\left({k_2{r}}\right)+{c_n}{Y_{n'}}\left({k_2{r}}\right)\right]e^{in\phi},
\end{align}
where $ Y_{n'} $ is a Bessel function of the second kind with order $ n' $ so that $ n' $ is equal to $ n'=n\sqrt{\mu_t/\mu_r} $ and $ k_2=k_0\sqrt{\varepsilon_z\mu_t} $~\cite{Chen1}.

$ a_n $, $ b_n $, $ c_n $ and $ d_n $ are the Lorentz-Mie coefficients and can be found using the boundary condition that is, continuity  of $ z $-component of electric field and $\phi$-component of magnetic field at the two boundaries $ r=R_1 $ and $ r=R $.
If we write the boundary condition in matrix form for the boundary at $ r=R $ we have 
\begin{eqnarray}\label{7q}
D_n^H\left(R\right).\left[\begin{matrix}
{1}\\{a_n}\end{matrix}\right]&=&D_{n',2}^Y\left(R\right).\left[\begin{matrix}
{b_n}\\{c_n}\end{matrix}\right]
\end{eqnarray}
and for the boundary condition at $ r=R_1 $ we have
\begin{eqnarray}\label{8q}
D_{n,1}^Y\left(R_1\right). \left[\begin{matrix}
{d_n}\\{0}\end{matrix}\right]=D_{n',2}^Y\left(R_1\right). \left[\begin{matrix}
{b_n}\\{c_n}\end{matrix}\right]
\end{eqnarray}\label{9q}
where $ D_n^{H}(R) $ and $ D_{n',m}^{Y}(m=1,2) $ are $ 2\times2 $ matrix and define as follow 
\begin{eqnarray}\label{10q}
D_n^{H}(R)=\left[\begin{matrix}{J_n\left(kR\right)}&{H_n^{(1)}}\left(kR\right)\\N{J'_n\left(kR\right)}&N{{H'}_n^{(1)}}\left(kR\right)\end{matrix}\right],
\end{eqnarray}
and
\begin{eqnarray}
D_{n',m}^{Y}(R_i)=\left[\begin{matrix}{J_{n'}\left({k_m}R_i\right)}&{Y_{n'}\left({k_m}R_i\right)}\\N_m{J'_{n'}\left({k_m}R_i\right)}&N_m{Y'_{n'}\left({k_m}R_i\right)}\end{matrix}\right].
\end{eqnarray}
In above equations $ N=\sqrt{\varepsilon_0\mu_0} $, $ N_2=\sqrt{\varepsilon_z\mu_t} $ and $ N_1=\sqrt{\varepsilon_c\mu_c} $.
Finding $\left[\begin{matrix}{b_n}\\{c_n}\end{matrix}\right]$ from equation (\ref{8q}) and substituting to equation (\ref{7q}) lead to 
\begin{eqnarray*}
\left[\begin{matrix}
{1}\\{a_n}\end{matrix}\right]=M_n.\left[\begin{matrix}
{d_n}\\{0}\end{matrix}\right],
\end{eqnarray*} 
where $ M_n $ is a $ 2\times2 $ matrix and can be found as follow
\begin{eqnarray*}
M_n=\left[\begin{matrix}{M_{n,11}}&{M_{n,12}}\\{M_{n,21}}&{M_{n,22}}\end{matrix}\right];
\end{eqnarray*}
and
\begin{eqnarray*}
M_n=\left[D_n^H\left(R\right)\right]^{-1}.{D_{n',2}^Y}\left(R\right).\left[D_{n',2}^Y\left(R_1\right)\right]^{-1}.{D_{n,1}^Y}\left(R_1\right).
\end{eqnarray*}

For the $ a_n $ and $d_n$ coefficients we have 
\begin{eqnarray}\label{an,dn}
d_n=\frac{1}{M_{n,11}}\\
{a_n}=\frac{M_{n,21}}{M_{n,11}},
\end{eqnarray}
where the scattering efficiency of the structure can be found using the $a_n$ coefficient as follow \cite{Bohren}
\begin{align}\label{11q}
{Q}=\frac{2}{kR}{\sum_{n=-\infty}^{+\infty}{\left| {{a_n}} \right|}^2}.
\end{align}
According to above equation to achieve the invisibility condition and minimizing the scattering cross section $a_n$ coefficient should be minimized. It is good to note that the resonance condition can be found by maximizing the $a_n$ coefficient.

For the $b_n$ and $c_n $ coefficients we have 
\begin{align}
b_n=\frac{P_{n,11}}{M_{n,11}},\\
c_n=\frac{P_{n,21}}{M_{n,11}};
\end{align}
where in $P_n$ is a $2 \times 2$ matrix and can be calculated using eqs.(\ref{7q}), (\ref{8q}) and (\ref{an,dn}) as follow
\begin{align}
P_n=\left[\begin{array}{cc}
P_{n,11}&P_{n,12}\\
P_{n,21}&P_{n,22}
\end{array}\right];\\
P_n=\left[D_{n',2}^Y\left(R_1\right)\right]^{-1}.{D_{n,1}^Y}\left(R_1\right).
\end{align}

\section{Hyperbolic Layered Metamaterial} \label{ENZ}
\begin{figure}[]
\center
\includegraphics[width=\linewidth]{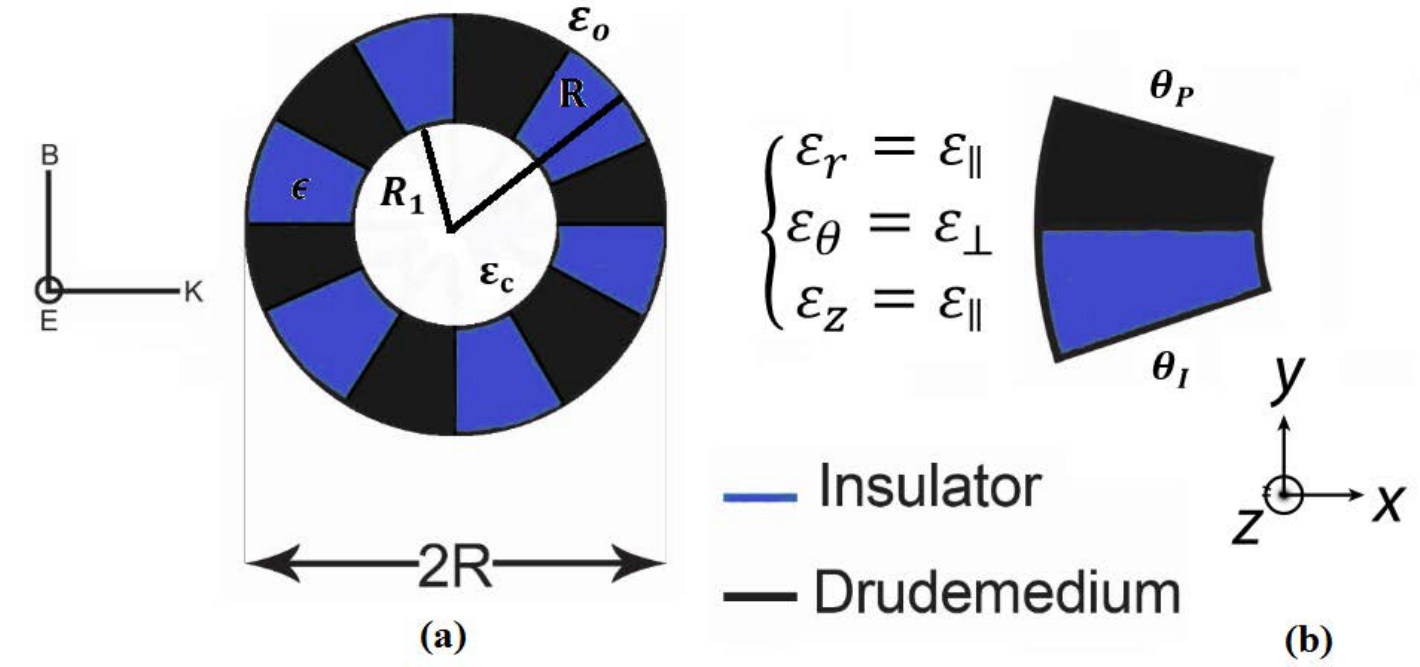}
\caption{(a) Structure of a proposed layered metamaterial for TE illumination.  (b) The unit cell of the structure in part (a) composed of a layer of Drude medium and an insulator layer.}\label{metamaterial}
\end{figure}

To design an anisotropic material so that scattering suppression can happen; we use the feature of layered metamaterial. So we consider a composite structure of two materials as we show in Fig.\ref{metamaterial}.

The unit cell of the structure is a metal layer with angle thickness $ \theta_P $ and a dielectric medium with angle thickness $ \theta_I $. Thus the filling factor for the metal medium can be expressed by 
\begin{align*}
f=\frac{\theta_P}{\theta_P+\theta_I}.
\end{align*}
we set the permittivity of the dielectric medium $ \varepsilon_I=10 $ as in the ref \cite{carlos}.
For the plasmonic medium we apply the Drude model as follow \cite{Maier}
\begin{align}\label{12q}
{\varepsilon_P}(\omega)=1-\frac{\omega_p^2}{\omega^2-i\gamma\omega}
\end{align}
where $ \omega_p $ is the plasma frequency of the metal and $ \gamma $ is damping factor. Since the real part of permittivity for the metal is negative, for low-loss metal $(\gamma<<\omega_p)$, $\omega$ should be less than $ \omega_p $ $(\omega<\omega_p)$, so throughout the paper for our calculation we work with the frequency less than plasma frequency.

\begin{figure*}[]
\center
\includegraphics[width=\linewidth]{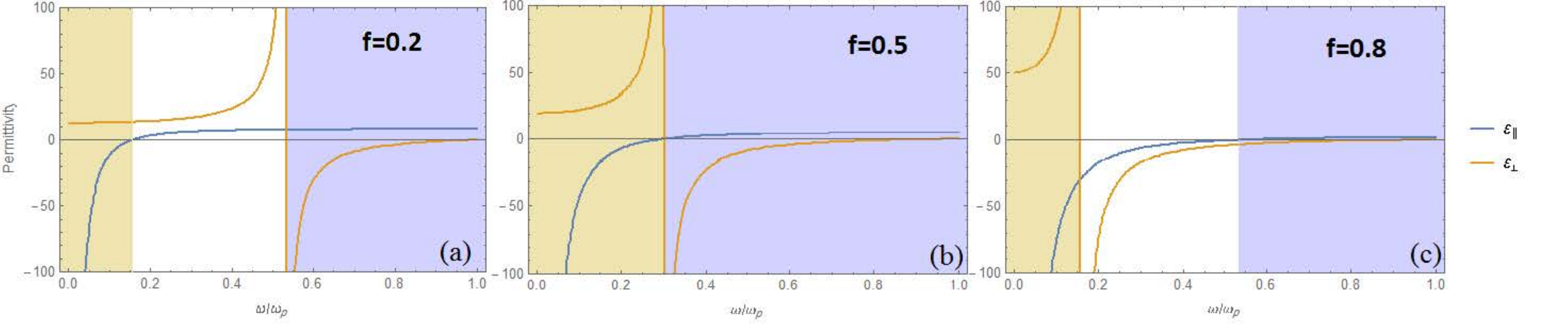}
\caption{Real part of the two components of effective permittivity $\epsilon_{\parallel}$(blue line) and $\epsilon_{\perp}$(orange line) in a layered metamaterial composed of a Drude medium with $\gamma=\omega_P/100$ and a dielectric layer with permittivity $\epsilon_I=10$ for three different filling factors (a) $f=0.2$, (b) $f=0.5$, (c) $f=0.8$. The shaded regions denote spectral bands where the metamaterial exhibits a hyperbolic dispersion of the type I (shaded in brown) and the type II (shaded in blue). }\label{effective}
\end{figure*}

Inasmuch as the permeability of the plasmonic and dielectric media is equal to one $\mu_p=\mu_I=1$ so the effective permeability of the composite structure would be $ \mu=1 $. Hence, for our calculation we have $n'=n$.  On the other hand, effective medium theory give the following relations for the effective permittivity of the metamaterials in parallel and perpendicular directions \cite{EMT}
\begin{align}\label{13q}
\varepsilon_\parallel=f\varepsilon_P+(1-f)\varepsilon_I
\end{align}
and
\begin{align}\label{14q}
\varepsilon_\perp=\frac{\varepsilon_I\varepsilon_P(\omega)}{f\varepsilon_I+(1-f)\varepsilon_P(\omega)}
\end{align}
where perpendicular and parallel is referred to the directions of the electric field with respect to the layered intersection.
A glance view to the above equation shows that $ \varepsilon_{\parallel} $ goes to zero in some frequencies which are obtained from the relation \cite{carlos}
\begin{align}\label{15q}
\omega_{zero}=\frac{\sqrt{(\omega_p^2-\gamma^2)f-\gamma^2(1-f)\varepsilon_I}}{\sqrt{\varepsilon_I-f(\varepsilon_I-1)}}.
\end{align}
Fig.\ref{effective} shows the variation of the real parts of the components of effective permittivity in the structure versus frequency for three different filling factors $f=0.2$ (part (a)), $f=0.5$ (part (b)) and $f=0.8$ (part (c)). The regions in which Re$\epsilon_{\parallel}$Re$\epsilon_{\perp}<0$  have  hyperbolic dispersion and are shaded in the pictures. The type I(II) hyperbolic media with Re$\epsilon_{\parallel}<0$(Re$\epsilon_{\perp}<0$) are shaded in brown(blue)\cite{hyperbolic1,hyperbolic2}.

\section{Result and Discussion}\label{result}
In order to find the condition in which the invisibility is taken place we analyse the scattering cross section according to the relation (\ref{11q}) for different filling factor in different frequency. we truncate the summation in eq.(\ref{11q}) to $ n=1 $. Therefore, we just consider dipole approximation to the scattering efficiency and the contribution of higher order is ignored which requires the small dimension of the cylinder compare to wavelength \cite{Bohren}.

The diagram in Fig.\ref{Q-R1-T-0.5}, shows the scattering efficiency in different frequency according to eq. (\ref{11q}) for the structure of Fig. \ref{metamaterial}  for different filling factor. This scattering spectrum is evaluated with assuming a low-loss Drude medium with $ \gamma={\omega_p}/{100} $. The anisotropic cavity is embedded in air with $ \varepsilon_0=\varepsilon_c=1 $ and it is small enough ($ R_1=T=k_P^{-1}/2 $ where $ k_P=\omega_P/c $) so that the series truncation would be valid.

\begin{figure}[]
\center
\includegraphics[width=\linewidth]{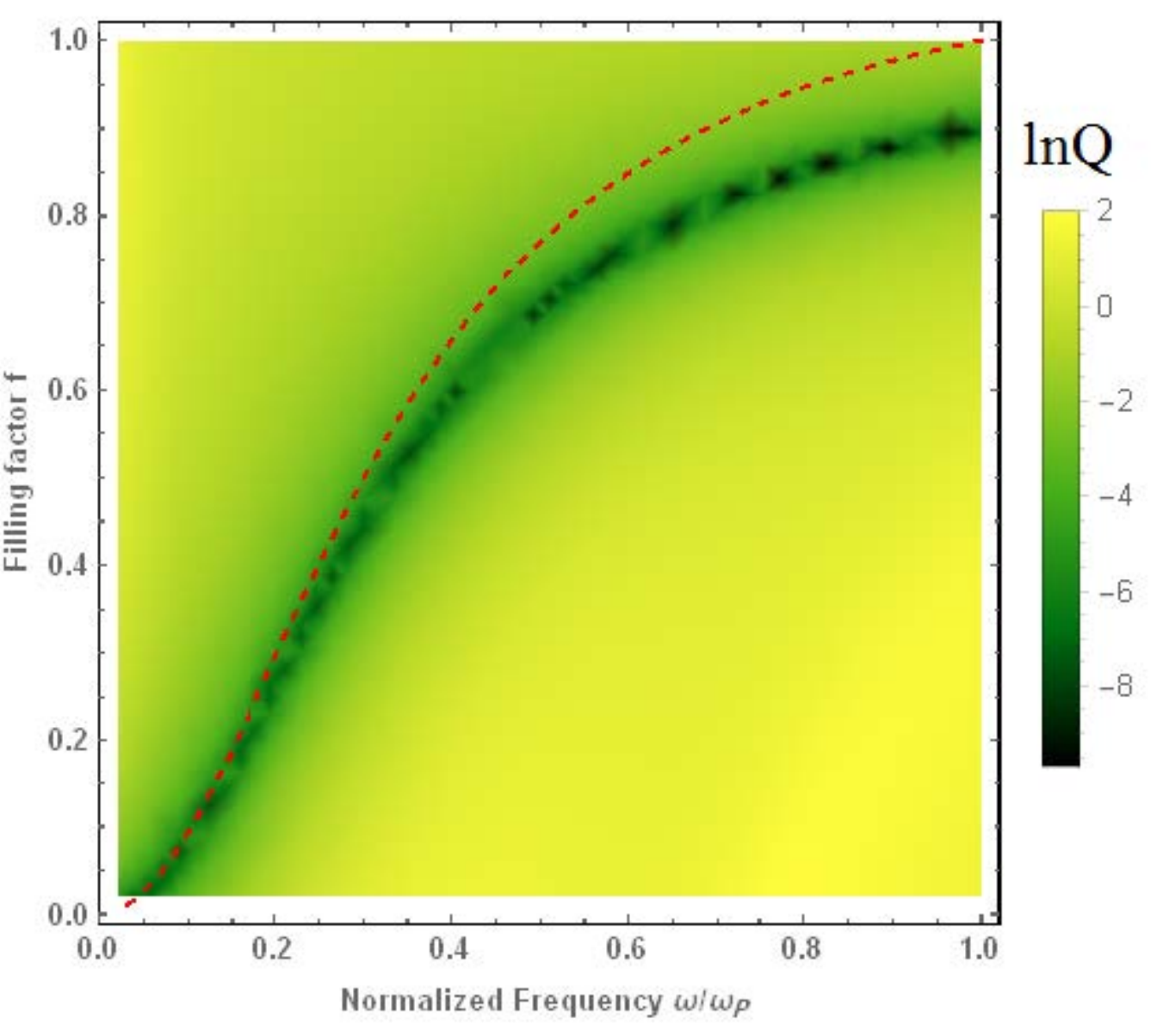}
\caption{Scattering cross section in logarithmic scale for a cylinder cavity with $R1=T=k_p^{-1}/2$ immersed in air $\epsilon_0=\epsilon_c=1$. The damping coefficient of the Drude medium is $\omega_P/100$ and the filling factor varies from $0$ to $1$. The Red dashed line represent the $\omega_{zero}$ in eq. (\ref{15q}).}\label{Q-R1=T=0.5}
\end{figure}

We observe a curve with dark color which shows the region with scattering cancellation. In Fig.\ref{Q-R1=T=0.5} we also show the $ \omega_{zero} $ curve and as an interesting result we observe that the $ \omega_{zero} $ curve have a very good coincidence with scattering cancellation which means that scattering cancellation can happen in the frequency in which the effective permittivity of the structure $ \varepsilon_\parallel $ goes to zero. Such outcome has been previously predicted by Kim et. al. in ref. \cite{kim} and is also in agreement with the result reported in ref.~\cite{carlos} for TM polarization.

The frequency in which the scattering cancellation (invisibility frequency) is happened can be tuned by changing the filling fraction of the Drude medium $f$ in the whole structure. One can see that with increasing the energy and getting close to plasma frequency $ \omega_P $; the two curves (the dark and the red-dashed curves) getting away of each other and so the Kim's approach is no longer valid. It may be duo to the effective medium theory eqs.(\ref{13q}) and (\ref{14q}) lose its validation in small wavelength \cite{EMT}. This result is explained in ref.~\cite{carlos} for TM polarization as well.

\begin{figure*}[]
\center
\includegraphics[width=\linewidth]{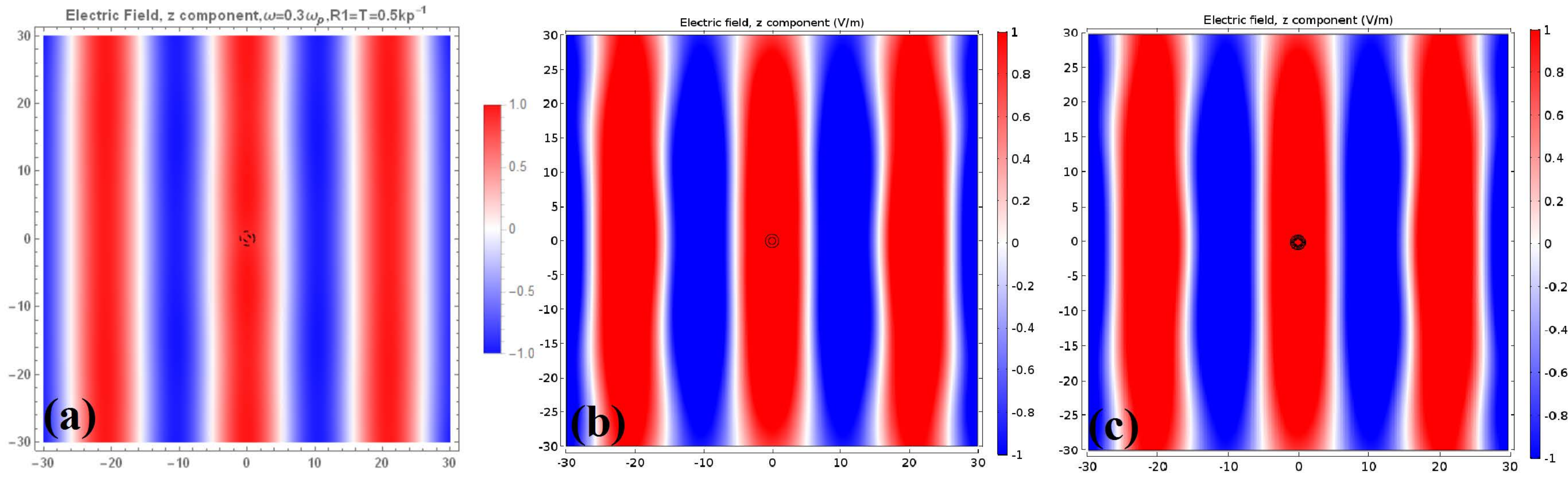}
\caption{Scattering pattern for the electric field $z$-component of the wave which is obtained based on (a) analytical approach (b) and (c) finite element method. The scatterer in parts (a) and (b) is an anisotropic  cavity cylinder with the size $R_1=T=k_p^{-1}/2$ and with zero permittivity in $z$ direction and in part (c) is a layered metamaterial with the structured shown in Fig.~\ref{metamaterial} with effective zero permittivity. The operating frequency is $\omega=0.3\omega_P$ and the filling factor for Drude medium in part(c) is $f=0.5$.}\label{com-R1=t=0.5}
\end{figure*}

To show that invisibility is really happened we represent in Fig.~\ref{com-R1=t=0.5} the scattering pattern for the $z$-component of electric field in passing through the cylinder cavity. Two approaches are applied to create the electric field diagrams in Fig.~\ref{com-R1=t=0.5}. In part (a) the analytical approach using the relations (\ref{2q}), (\ref{5q}) and (\ref{6q}) and in part (b) the finite element solution using COMSOL Multiphysics is used. The feature of the scatterer including the size and constitutive parameter in Fig.~\ref{com-R1=t=0.5} is exactly the same as the one in Fig.~\ref{Q-R1=T=0.5}. It is worth noting that hereafter, in producing the scattering pattern with analytical approach, like in Fig.~\ref{com-R1=t=0.5}(a), we truncate the series for electric field in eqs.(\ref{2q}), (\ref{5q}) and  (\ref{6q}) to $N_{max}=15$. We also bring in part (c) a finite element simulation for a layered metamaterial with the structure shown in Fig.~\ref{metamaterial} composed of ten unit cells in azimuthal direction.  First of all the figure shows that in all three parts of the figure the wave fronts are kept in passing through the ENZ cylinder. Second, it is clear that the scattering pattern for the layered metamaterial composed of a plasmonic medium and a dielectric in part (c) has a very good agreement with the one in part (b) which is an anisotropic cavity with zero permittivity in $z$ direction. This shows that the layered structure in part (c)  has indeed effective zero permittivity. We choose the filling factor to be $f=0.5$, so the scattering cancellation based on the result in Fig.~\ref{Q-R1=T=0.5} is happened in $\omega=0.3 \omega_P$.

\begin{figure}[]
\center
\includegraphics[width=\linewidth]{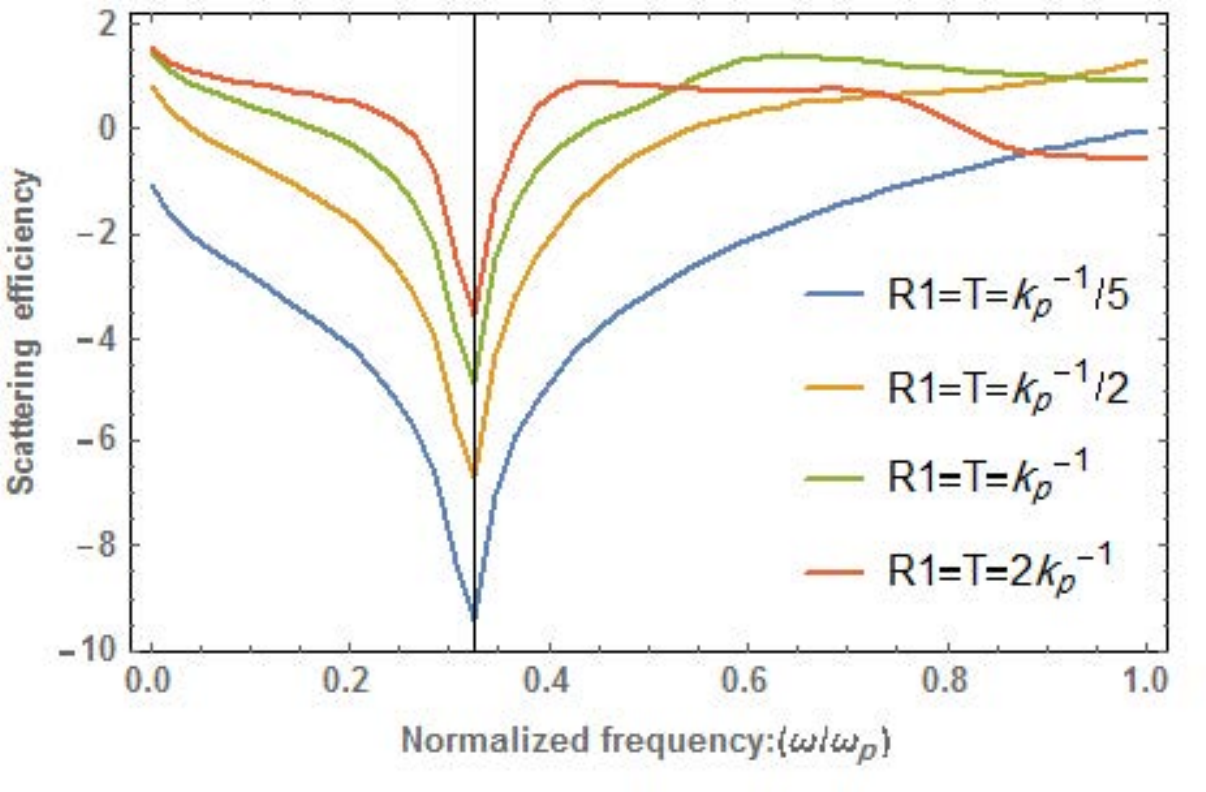}
\caption{Scattering cross section in logarithmic scale for cylinder cavities with the sizes $T=k_P^{-1}/5 $, $ T=k_P^{-1}/2 $, $ k_P^{-1} $ and $2 k_P^{-1} $ immersed in air $\epsilon_0=\epsilon_c=1$. The damping coefficient of the Drude medium is $\omega_P/100$ and the filling factor choose to be $f=0.5$. The frequency of invisibility for all scatterers are $\omega=0.3\omega_P$. }\label{compare}
\end{figure}

\begin{figure*}[]
\begin{minipage}{0.23\linewidth}
\caption{The diagram for scattering cross section in logarithmic scale for cylinder cavities with the size  (a) $R_1=T=k_P^{-1} $, and (b) $ R_1=T=2k_P^{-1} $, immersed in air $\epsilon_0=\epsilon_c=1$. The damping coefficient of the Drude medium is $\omega_P/100$. }\label{Q2}
\end{minipage}
\begin{minipage}{0.75\linewidth}
\center
\includegraphics[width=\linewidth]{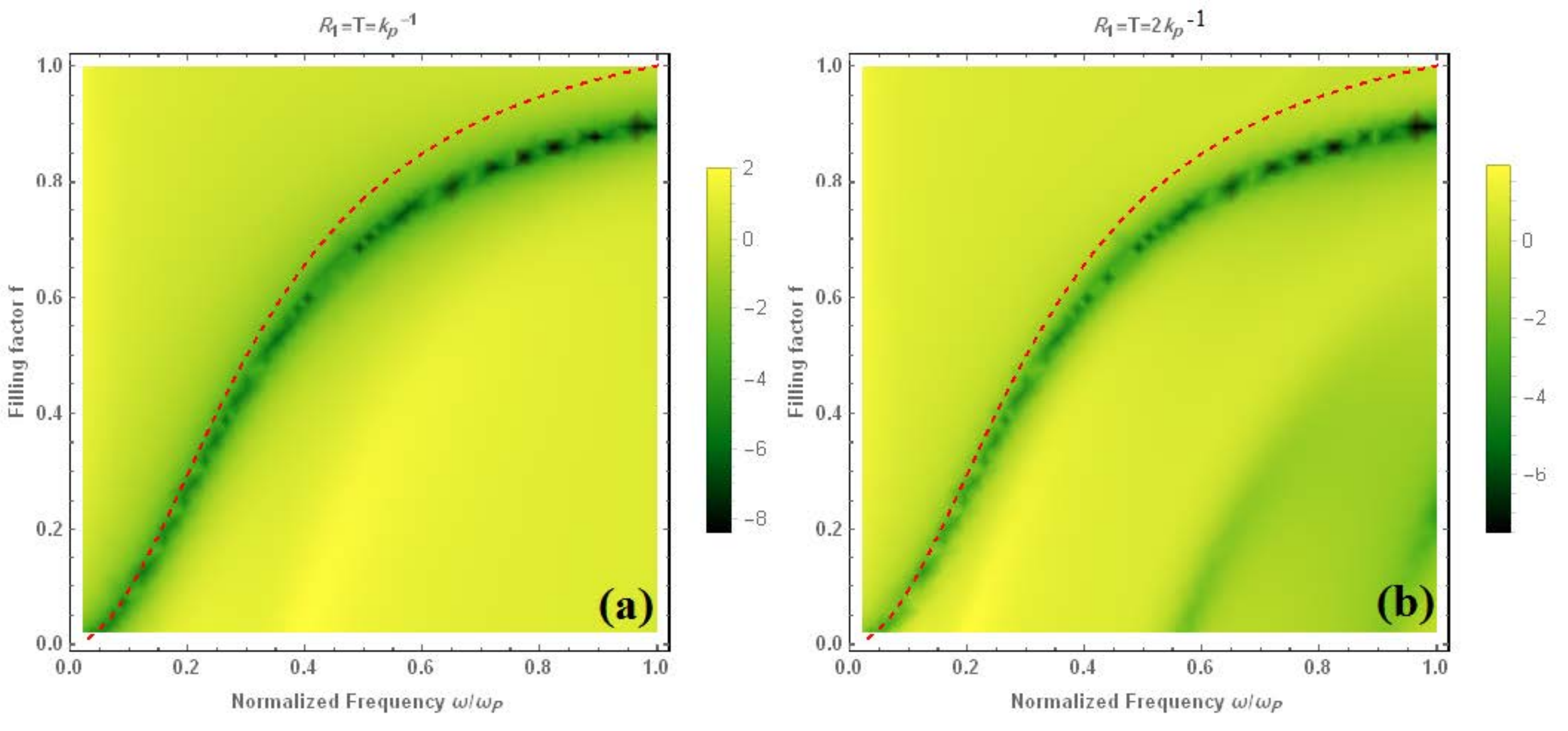}
\end{minipage}
\end{figure*}

For comparison and to find the criteria for dipole approximation we increase the size of the scatterer and plot the scattering efficiency  diagram for four different  situations with $T=k_P^{-1}/5 $, $ k_P^{-1}/2 $, $ k_P^{-1} $ and $2 k_P^{-1} $ so that $ T/R_1=1 $ is conserved. Again we set $ \gamma=\omega_P/100 $ and here $ f=1/2 $. As we see from Fig.~\ref{compare}, the least scattering efficiency is for $ T=k_P^{-1}/5 $, the smallest scatterer which is completely acceptable. The invisibility frequencies for all the scatterers are the same and are set at about $\omega\simeq\omega_{zero}=0.3\omega_P$; however, the scattering efficiency grows significantly with increasing the size of the scatterer. One can see the scattering efficiency diagram for scatterer with the size $R_1=T=k_p^{-1}$ and $R_1=T=2k_p^{-1}$ in Fig.~\ref{Q2} part (a) and (b), respectively.

\begin{figure*}[]
\center
\includegraphics[width=\linewidth]{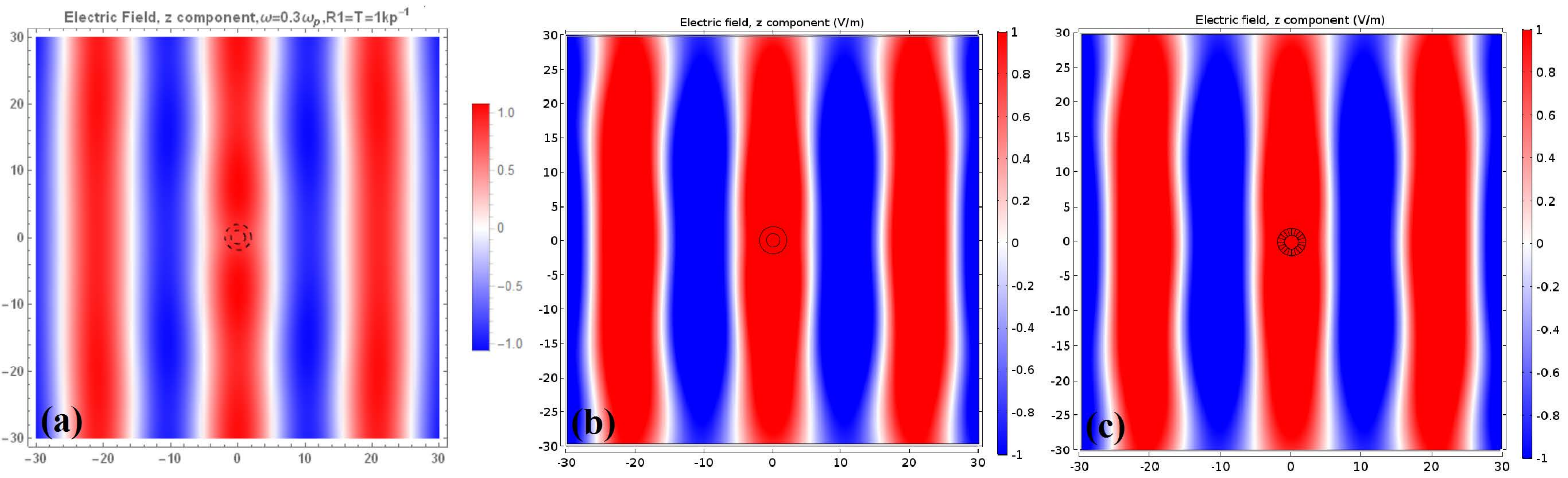}
\caption{Scattering pattern for the electric field $z$-component of the wave which is obtained based on (a) analytical approach (b) and (c) finite element method. The scatterer in parts (a) and (b) is an anisotropic cavity cylinder with the size $R_1=T=k_P^{-1}$ and zero permittivity in $z$ direction and in part (c) is a layered metamaterial with the structured shown in Fig.~\ref{metamaterial} with effective zero permittivity. The operating frequency is $\omega=0.3\omega_P$ and the filling factor for Drude medium in part(c) is $f=0.5$. }\label{com-R1=t=1}
\end{figure*}

\begin{figure*}[]
\begin{minipage}{0.33\linewidth}
\caption{Scattering pattern for the electric field $z$-component of the wave which is obtained based on (a) analytical approach (b), (c) and (d) finite element method. The scatterer in parts (a) and (b) is an anisotropic cavity cylinder with the size $R_1=T=2k_p^{-1}$ and zero permittivity in $z$ direction, in part (c) is a layered metamaterial with the structured shown in Fig.~\ref{metamaterial} with effective zero permittivity and in part (d) the scatterer is an isotropic cylindrical shell with permittivity $\varepsilon=5$. The operating frequency is $\omega=0.3\omega_P$ and the filling factor for Drude medium in part(c) is $f=0.5$.}\label{com-R1=t=2}
\end{minipage}
\begin{minipage}{0.66\linewidth}
\center
\includegraphics[width=\linewidth]{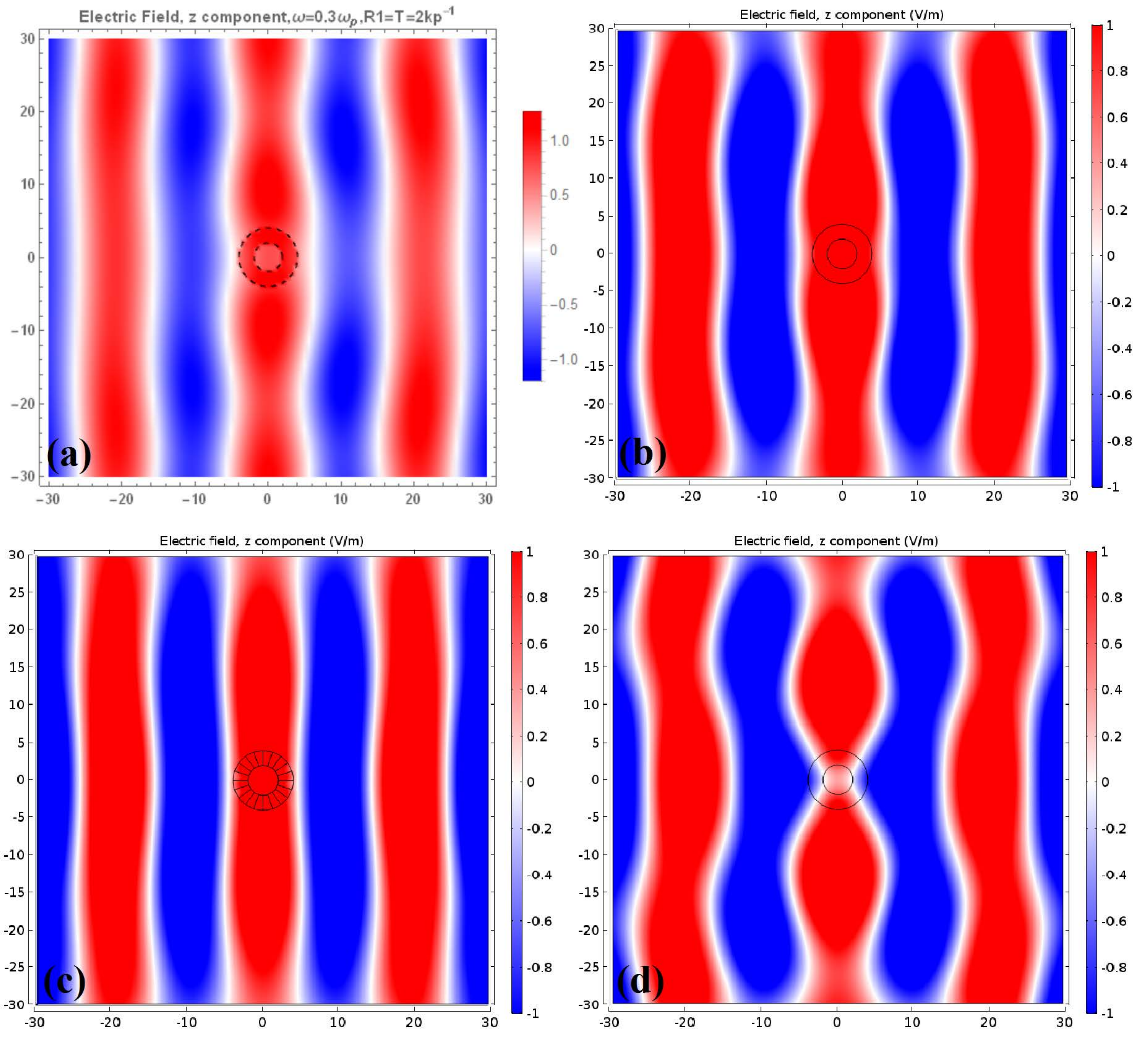}
\end{minipage}
\end{figure*}

The scattering pattern for the $z$-component of electric fields in the both cases which is explained in Fig.~\ref{Q2} (a) and (b) are also brought in Fig.~\ref{com-R1=t=1} and \ref{com-R1=t=2} respectively. The patterns are obtained again using two approaches, analytical solution part (a)s as well as finite element method part (b) and (c)s. 

In comparing Figs.~\ref{com-R1=t=0.5}, \ref{com-R1=t=1} and \ref{com-R1=t=2} parts (a)-(c), it is clear that with increasing the size of the scatterer the distortion of the wave front of the incident wave is increased when we making the size of the cylinder larger. In Fig.~\ref{com-R1=t=2}(d) we also bring a isotropic cylinder cavity with the relative permittivity $\varepsilon=5$ for the shell layer and in the same size with the scatterer in Fig.~\ref{com-R1=t=2}(a)-(c) to show the role of ENZ medium in cancellation of scattering cross section. The remarkable distortion of the wave front in part (d) of the Fig.~\ref{com-R1=t=2} in comparison with part (a)-(c) approve this fact.

\begin{figure*}
\center
\includegraphics[width=\linewidth]{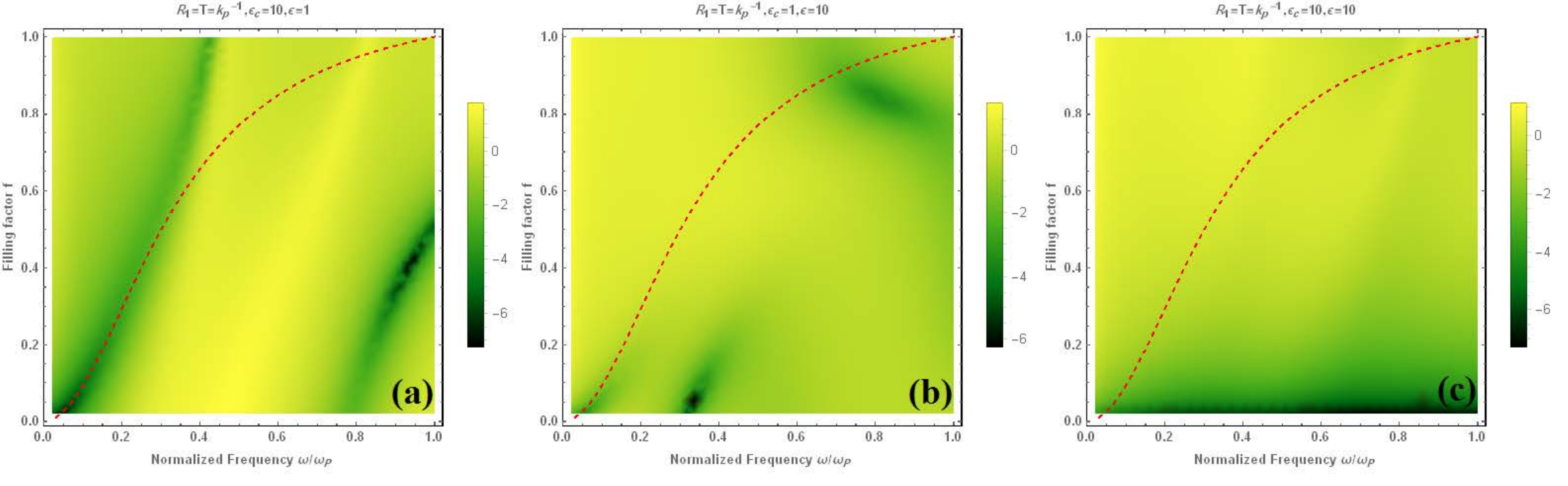}
\caption{Scattering efficiency diagram for a cylindrical cavity with zero effective permittivity immersed in a medium with core medium permittivity $\varepsilon_c$ and outside medium permittivity $\varepsilon_0$. (a) $\varepsilon_c=10$, $\varepsilon_0=1$ (b) $\varepsilon_c=1$, $\varepsilon_0=10$ and (c) $\varepsilon_c=10$, $\varepsilon_0=10$. The size of the cylinder is chosen to be $R_1=T=k_p^{-1}$.}\label{Q-epsilon}
\end{figure*}

Another parameter which we want to analysed is the role of the relative permittivity of the medium which the ENZ cylinder shell immersed in. So in Fig.~\ref{Q-epsilon} the scattering efficiency of an ENZ cylinder cavity with the size $R_1=T=k_p^{-1}$ in three different situations is shown. In Fig.~\ref{Q-epsilon}(a) the relative permittivity of the core and external medium is chosen to be $\varepsilon_c=10$, $\varepsilon_0=1$, respectively. As is clear the curve of scattering cancellation is quite different with the situation in which $\varepsilon_c=1$, see Fig.~\ref{Q2}(a), however, it still varies near the $\omega_{zero}$ curve specially for the lower energies. There is also an extra scattering cancellation curve in comparing with Fig.~\ref{Q2}(a) for higher energy which is far from the $\omega_{zero}$ curve. In Fig.~\ref{Q-epsilon}(b) and (c) the relative permittivity for the core and outside medium is chosen to be $\varepsilon_c=1$, $\varepsilon_0=10$ and $\varepsilon_c=10$, $\varepsilon_0=10$, respectively. One can see that there is almost no curve for scattering suppression specially for part (c). The result extracted from Fig.~\ref{Q-epsilon}(a)-(c) express that the frequency for scattering cancellation is depend on the permittivities of the core and environment medium and can be different from the $\omega_{zero}$ curves. These result may be in disagreement with the Kim's prediction \cite{kim} and cause to decrease the importance of the condition given by $\omega_{zero}$ as reported in ref.~\cite{carlos}.

At last it is good to note that unlike to the result represented in ref.~\cite{carlos}, in all the diagrams for the scattering efficiency in Figs.~\ref{Q-R1=T=0.5}, \ref{Q2} and \ref{Q-epsilon} there is no curve for the resonance condition which means that no plasmonic excitation is happened for the structure shown in Fig.~\ref{metamaterial} with the assumed polarization. The reason behind this differences is that here in our structure with the polarization which we choose for the light, the electric field of the light has no normal component to the metal layers in the layered metamaterial structured shown in Fig.~\ref{metamaterial} and the electric field of the incident wave is tangential to all the plasmonic layers in the structure. Therefore, plasmonic resonance could not happen and there is no maximum for scattering cross section.
\section{conclusion}
In conclusion, we studied the scattering behavior of the uniaxial cylindrical cavity analytically. Then we proposed a layered metamaterial structure to create the above uniaxial anisotropic structure practically so that it prepared the hyperbolic material with effective required constitutive parameter in all directions. 

The scattering properties of this metamaterial was analysed under different filling factor with changing the frequency. We found out that the curve of scattering cancelation has a very good agreement with the $\omega_{zero}$ curve which shows the frequency points for different filling factor in which the effective permittivity of the metamaterial structure goes to zero. This result has a good conformity with the Kim's prediction. But when we changed the relative permittivities of the environmental media the coincidence of the two curves was removed. This result showed that the scattering cancellation not only depend on permittivities of the scatterer but also depends on the relative permittivities of the environment medium of the scatterer which might not be in agreement with Kim's prediction.

The results extracted with analytical solution is accompanied with some finite element simulations for the electric field of the wave to approve that the invisibility condition is indeed acceptable.


\end{document}